\documentclass[preprint,authoryear,2p,review,12pt]{elsarticle}
\usepackage{lineno}\linenumbers*[1]

\usepackage{epsfig}

\usepackage{amssymb}

\biboptions{square}

\journal{Journal of Atmosphere and Solar-Terrestrial Physics}

\begin{document}

\begin{frontmatter}



\title{Dependence of GCRs influx on the Solar North-South Asymmetry}


\author[label1,label2]{Il-Hyun Cho} 
\author[label1]{Young-Sil Kwak} 
\author[label3]{Heon-Young Chang\corref{cor1}}\ead{hyc@knu.ac.kr}
\cortext[cor1]{corresponding author}
\author[label1]{Kyung-Suk Cho} 
\author[label1]{Young-Deuk Park} 
\author[label1,label2]{and Ho-Sung Choi} 

\address[label1]{Korea Astronomy and Space Science Institute, Daejeon, 305-348, Korea}
\address[label2]{University of Science and Technology, Daejeon, 305-348, Korea}
\address[label3]{Department of Astronomy and Atmospheric Sciences, Kyungpook National University, Daegu, 702-701, Korea}

\begin{abstract}
We investigate the dependence of the amount of the observed galactic cosmic ray (GCR) influx on the solar North-South asymmetry using the neutron count rates obtained from four stations and sunspot data in archives spanning six solar cycles from 1953 to 2008. We find that the observed GCR influxes at Moscow, Kiel, Climax and Huancayo stations are more suppressed when the solar activity in the southern hemisphere is dominant compared with when the solar activity in the northern hemisphere is dominant.  Its reduction rates at four stations are all larger than those of the suppression due to other factors including the solar polarity effect on the GCR influx. We perform the student's t-test to see how significant these suppressions are. It is found that suppressions due to the solar North-South asymmetry as well as the solar polarity are significant and yet the suppressions associated with the former are larger and more significant.
\end{abstract}


\end{frontmatter}

\linenumbers

\section{Introduction}
It is well known that the level of galactic cosmic ray (GCR) influx to the terrestrial atmosphere is variable in time. The observed  GCR influx is anti-correlated with the observed sunspot number[\citealt{11,49}]. The higher level of the solar activity induces a stronger magnetic field strength in the heliosphere, and consequently reduces the amount of GCRs observed in the vicinity of the Earth. For instance, during the increasing phase of the solar cycle, the  observed level of GCRs near the Earth has been gradually reduced. A long-term effect is also studied by Lockwood et al. (1999) using several geo- and solar-magnetic indices from 1964 to 1995 corresponding to solar cycles from 20 to 22. They showed that measurements of the near-Earth interplanetary magnetic field reveal that the total magnetic flux leaving the Sun has risen by a factor of 1.4 since 1964, which interestingly reflects the increase of solar source magnetic flux, which also indirectly implies the decrease of GCR influx.

In addition to the dependence of the GCR flux on the strength of the solar activity, the heliospheric current sheet is asymmetric with respect to the heliographic equatorial plane [\citealt{21,17,31}]. For instance, a southward offset of the current sheet has been reported while ULYSSES monitored the cosmic-ray proton intensity from September 1994 to May 1995[\citealt{43,42}]. We note that the observing epoch corresponds to the period when the solar southern hemisphere is magnetically dominant. As one of the solar northern- and southern-hemispheres is dominant in a given solar activity, the magnetic field of the heliosphere near the Sun is likely to follow a similar behavior. \Citet{50}  have also reported that the electron temperature and magnetic field in  the solar southern coronal hole are higher and stronger than those in the northern coronal hole during the period from 1992 to 1997. This is also the time interval when the solar activity in the southern hemisphere is dominant, as we mentioned.

In this paper, we investigate whether the level of the GCR influx shows any dependence associated with the solar North-South asymmetry using the neutron count rates obtained from 4 stations and sunspot data archived spanning six solar cycles from 1953 to 2008. This is an interesting question not only in itself but also because of the following reason. According to what is mentioned above one may expect that the average location of the current sheet around the Earth can be quasi-periodically varying with the solar North-South asymmetry. Thus, the observed GCR influx may be changing in magnitude not only with the solar cycle but also with the solar North-South asymmetry.

\section{Data and Results}

The GCR data are recorded at ground-based neutron monitors, which detect variations in the low energy part of the primary cosmic ray spectrum. The lowest energy that can be detected  at the top of the atmosphere depends on the geomagnetic latitude, and ranges from 0.01 GeV at stations near the geomagnetic poles to about 15 GeV near the geomagnetic equator. The National Geophysical Data Center(NGDC) tabulates the daily average GCR count rate per hour for a number of stations. From the website\footnote{${\rm http://www.ngdc.noaa.gov/}$}, we have taken the yearly mean of the GCR influx  observed at 4 stations,  where GCR data sets having a time-span longer than 20 years in both prior and post to $\sim$ 1980 are available: Moscow($55^{\circ}$N, $37^{\circ}$E, 2.42GV), Kiel($54^{\circ}$N, $10^{\circ}$E, 2.32GV), Climax($39^{\circ}$N, $106^{\circ}$W, 2.99GV) and Huancayo($12^{\circ}$S, $75^{\circ}$W, 12.92GV).

We have adopted the yearly averaged wolf number (SN) from 1953 to 2008 provided by Solar Influences Data Analysis Center (SIDC)\footnote{${\rm http://sidc.oma.be/index.php3}$} for the GCR-SN plot. We have also taken sunspot area data for the calculation of the solar North-South asymmetry from the NASA website\footnote{${\rm http://solarscience.msfc.nasa.gov/greenwich.shtml}$}, which is regularly updated by Hathaway at the  Marshall Space Flight Center(MSFC). Text files are there available containing the yearly averages of the daily sunspot areas (in units of millionths of a hemisphere) in the northern and solar southern hemispheres separately.

In Figure 1 we show the running average of the solar North-South asymmetry, which is defined as the difference of the sunspot area between solar northern and southern hemispheres normalized by their sum, $(A_{N}-A_{S})/(A_{N}+A_{S})$, for the period of $1953 - 2008$ as a function of time. In this plot the length of the moving window is 7 years. When values of the solar North-South asymmetry is positive the northern solar hemisphere is more active, and vice versa. Note that the dominance of the active hemisphere is reversed at $\sim$ 1980. That is, in general the solar northern and southern hemispheres are magnetically dominant before and after $\sim$ 1980, respectively.

In Figure 2, we show yearly GCR fluxes at four neutron monitor stations as a function of the sunspot number. We divide the observed GCR flux and SN data sets into two sub-sets,  respectively, separated at $\sim$ 1980 for the present analysis. Open and filled circles represent the observed GCR count rates in the sub-sets with time intervals before and after  $\sim$ 1980, respectively. Straight lines in each panel result from the least square fits of two sub-sets. It should be noted that the averaged value of the observed GCR count rates represented by filled circles is smaller than that represented by open circles while the slopes of the best fits in each panel are more or less the  same. The mean differences of  GCR count rates between two sub-sets in Moscow,  Kiel, Climax and Huancayo are  2.83$\pm$0.33\%, 2.50$\pm$0.44\%, 2.72$\pm$0.18\% and 0.72$\pm$0.04\%, respectively. What it means is that while the anticorrelation between the GCR flux and the sunspot number holds good regardless of observation durations, the averaged GCR flux is apparently more reduced when the solar southern hemisphere is more active than the northern hemisphere. The relative difference shown is calculated from the best fit. We compute student's t and its probability for each station to see how significantly two mean values differ from. Student's t values for Moscow, Kiel, Climax and Huancayo are 9.50, 7,38, 6.28 and 5.21. Their false-alarm levels are all less than 0.01. It should be noted that the percentage variations in absolute counting rates during the solar cycle are different for different geomagnetic latitudes. The effect is smaller at lower latitudes in agreement with the shielding effect of the Earth's magnetic field on high-energy charged particles. Note that the vertical rigidity cut-offs for Moscow, Kiel, Climax, and Huancayo are 2.42GV, 2.32GV, 2.99GV, and 12.92GV, respectively. As a result, the mean value of data points in each sub-set is considered as significantly different. This conclusion is found to be insensitive to the epoch when we cut the data set. We have repeated our analysis with a few more sub-sets separated at different times rather than $\sim$ 1980, and obtained similar results.

To compare our findings with results from other possible factors causing the suppression we have observed in the present work, we have carried on our analysis with the data sets divided by three other criteria. By doing so we may wish to demonstrate that the solar North-South asymmetry is a more critical factor in reducing the observed GCR flux than any other solar variables compared in this paper.

In Figure 3, we show the GCR influx observed at the Moscow station as a function of  the sunspot number, as an example. In panel (a), the data set is divided by the level of the solar activity. That is, open and filled circles represent the observed GCR count rates in the period of strong and weak solar cycles, respectively. Open circles belong to the cycles 19, 21 and 22. Filled circles belong to the cycles 20 and 23. Note that cycles 19 and 20 correspond to the period prior to $\sim$ 1980, and cycles 21, 22 and 23 to the period post to $\sim$  1980. The solar cycle is defined such that it begins and ends at the solar minimum. A clear difference in the mean value cannot be seen. In panel (b), we show the observed GCR influx as a function of the sunspot number with respect to the individual solar cycle. Different symbols represent different solar cycles as indicated by the number. Note that the solar cycle is defined such that it begins and ends at the solar maximum unlike the case in panel (a). Having picked up the cycles 20 and 23, and compared them with other solar cycles by averaging, the results can be contrasted with panel (a). By defining the solar cycle in this way one may also see the polarity effect. In panel (c) the polarity effect on the observed GCR influx is shown. Open and filled circles represent the opposite polarity. Two consecutive solar cycles have the opposite polarity. As reported earlier[\citealt{56,55,57}], we confirm that the amount of the observed GCR influx varies as the polarity of the solar magnetic field is reversed. For comparison in magnitude, in panel (d) we show the result from the solar North-South asymmetry. It should be noted that the solar North-South asymmetry effect on the GCR influx is larger than any other effects. That is, the values of $dI/I$ due to the solar North-South asymmetry and the polarity are 2.83$\pm$0.33 \% and 1.36$\pm$0.50 \%, respectively, as shown in Table 1. Similar conclusions can be drawn from other stations.

It is also interesting, for instance, to note that the mean difference in the observed GCR influx is very large in cycles 20 and 23. These are the solar cycles whose signs both in the solar polarity and in the solar asymmetry are simultaneously opposite. These two factors apparently compete with  each other so that they can be either amplified or compensated accordingly. Hence we conclude that both effects should be considered with due care, particularly when one wishes to employ one of these effects into a sophisticated algorithm to predict anything accurately.

In Table 1, to summarize results, we list the difference in mean values and results of the student's t-test for sub-sets divided according to the solar polarity and the solar North-South asymmetry. In the first and second columns, neutron monitor stations and the criteria in dividing the data sub-sets are shown. In the third column the relative difference in the mean GCR flux is given. In the fourth, fifth, and sixth columns, results of student's t-test are given. We conclude that the suppression associated with the solar North-South asymmetry is  larger and more significant than that with the solar polarity.  We only consider in the present study the average long-term effect of the solar North-South asymmetry on the GCR influx. Though the solar North-South asymmetry also has shorter-term  periodicities than we reported in this paper[e.g., \citealt{04,01,02,03}], those short-terms  may not be notable due to the other effect such as a fluctuating shape of current sheet.

\section{Summary and Discussion}
We investigate whether the level of the GCR influx shows any dependence associated with the solar North-South asymmetry using neutron count rate data  obtained from 4 stations and sunspot data archived spanning six solar cycles from 1953 to 2008. We have found that the GCR influx observed at 4 neutron monitor stations are all suppressed when the solar southern hemisphere is more active. The amount of suppressions are 2.83$\pm$0.33\% at Moscow, 2.50$\pm$0.44\% at Kiel, 2.72$\pm$0.18\% at Climax and 0.72$\pm$0.04\% at Huancayo. These values are larger than the solar polarity effect observed at 4 stations (1.36$\pm$0.50\% at Moscow, 1.67$\pm$0.57\% at Kiel, 2.22$\pm$0.89\% at Climax and 0.39$\pm$0.04\% at Huancayo). We compute student's t values and its probability for each case, and come to conclusions that the solar North-South asymmetry effect on the GCR influx is larger than any other effects including the solar polarity effect.

Finally, we would like to add some speculation of its possible implication on the terrestrial climate. As Svensmark and Friis-Christensen (1997) pointed out, the decrease in the observed GCR influx may cause the low terrestrial albedo through reducing low level cloud cover, and consequently lead to the higher heat content at the surface of the Earth. In a sense, therefore, the heat content at the surface of the Earth could be different from that deduced by the measured TSI alone[\citealt{16}]. We suspect that the net-radiative energy at the surface of the Earth should be determined by two parameters even in a simple climate model based on an assumption that greenhouse fraction is constant[e.g., \citealt{09}]. For instance, two parameters can be the total solar irradiance (TSI), which is modulated with the solar cycle[\citealt{38}], and the Earth's albedo, which Svensmark and Friis-Christensen (1997) show varying with the observed GCR influx modulated. Hence one may wish to re-examine the temperature anomaly as a function of solar activity including, for example, the solar North-South asymmetry since the correlation between the GCR influx and the sunspot number is subject to the sign of the solar asymmetry[\citealt{05,13,14,15}].

\section{Acknowledgments}
We appreciate Katsuhide Marubashi for constructive comments on the GCRs distribution in the heliosphere and Kate Connors for careful reading the manuscript. We thank the anonymous referees for constructive comments which improve the original version of the manuscript. This work was supported by the 'Development of Korean Space Weather Center' of KASI and KASI basic research funds. HYC was supported by Basic Science Research Program through the National Research Foundation of Korea(NRF) funded by the Ministry of Education, Science and Technology (2009-0071263).

\bibliographystyle{elsarticle-harv}



\newpage

\begin{figure}
\centering
\noindent\includegraphics[width=29pc]{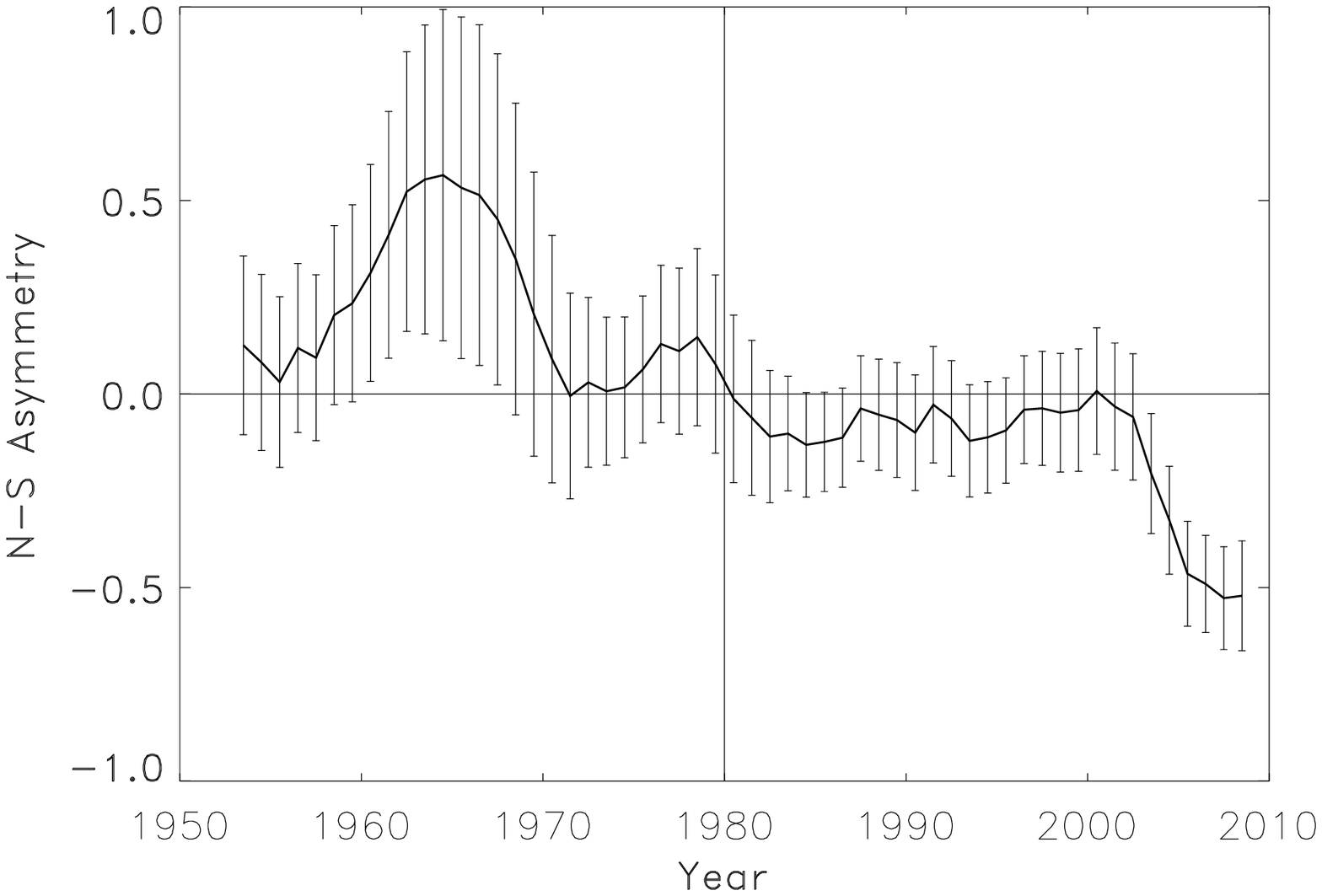} 
\caption{The running average of the solar North-South asymmetry with a 7-year window. The vertical line is given to separate the data set into two sub-sets. Note that the solar northern- and southern-hemisphere is seen more active prior to  $\sim$ 1980 and post to  $\sim$ 1980, respectively. Error bars are included.}
\end{figure}
\begin{figure}
\centering
\noindent\includegraphics[width=29pc]{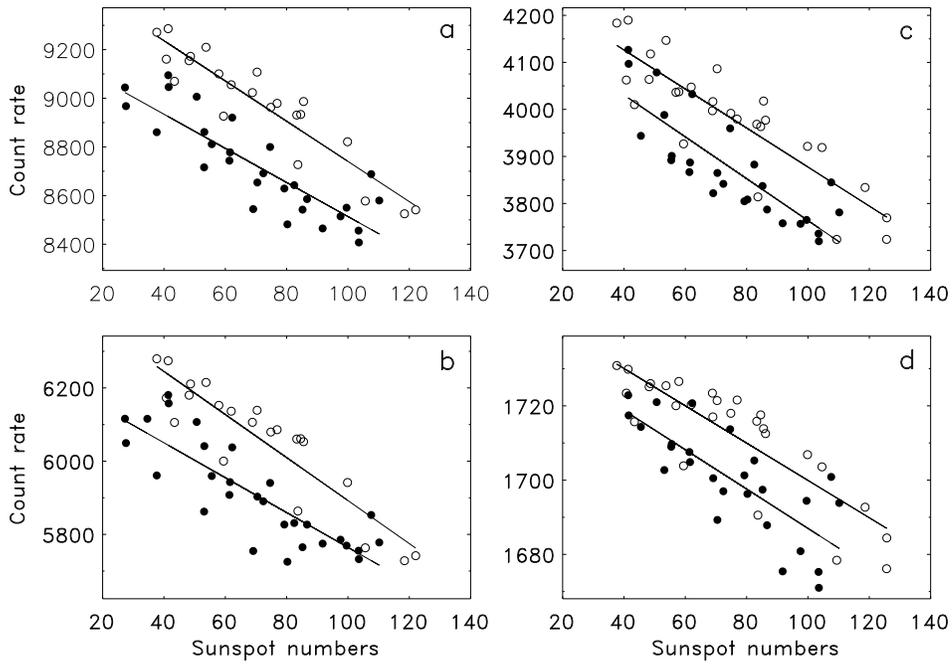} 
\caption{Observed GCR count rates measured at Moscow(a), Kiel(b), Climax(c) and Huancayo(d) as a function of sunspot numbers. Open and filled circles represent the obseved GCR influx in the period before and after 1980. Straight lines in each panel result from the least square fits of two sub-sets. }
\end{figure}
\begin{figure}
\centering
\noindent\includegraphics[width=29pc]{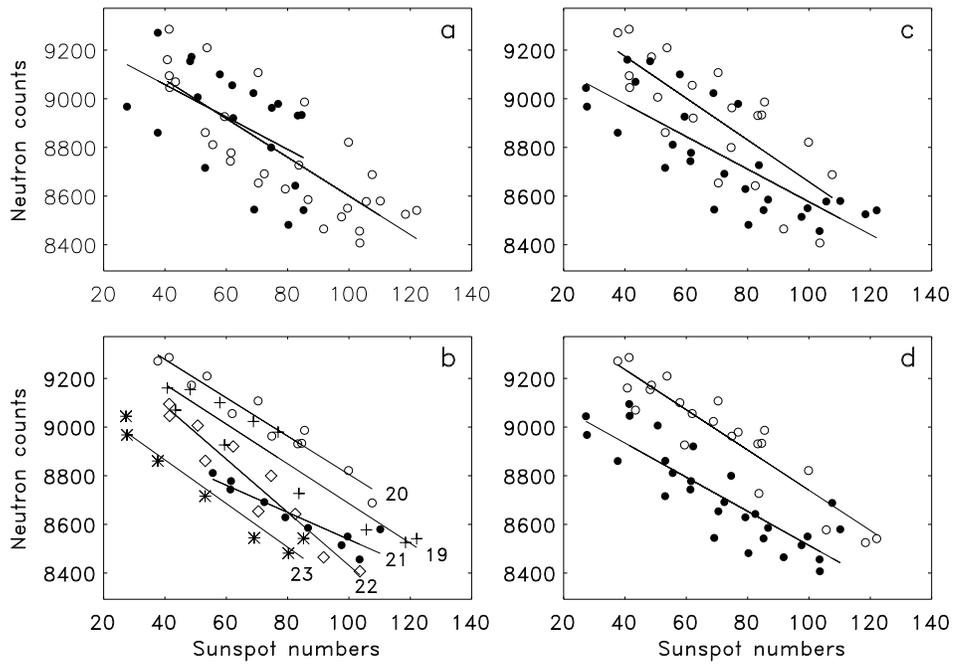} 
\caption{GCRs count rates measured at Moscow station as a function of sunspot numbers. Each straight line is the result of the least square fit. Different panels result from sub-sets divided by different criteria, i.e., (a) by the activity level of the solar cycles, (b) by the solar cycles, defined by the period between solar maxima,  (c) by the solar magnetic polarity, (d) by the solar North-South asymmetry.}
\end{figure}

\newpage

\begin{table}
\caption{Relative difference by the solar North-South asymmetry and polarity.}
\begin{tabular}{cccccc} \hline
Station & Factors & dI/I &Student-t&Deg. of freedom&Prob.(\%)  \\ \hline
Moscow & Polarity &1.36$\pm$0.50\%&2.83&48&$>$99.09 \\
 & Asymmetry &2.83$\pm$0.33\%& 9.50&48&$>$99.99 \\ \hline
Kiel & Polarity &1.67$\pm$0.57\%& 3.89&49&$>$99.95 \\
 & Asymmetry &2.50$\pm$0.44\%& 7.38&49&$>$99.99 \\ \hline
Climax & Polarity &2.22$\pm$0.89\%&4.78&51&$>$99.99 \\
 & Asymmetry &2.72$\pm$0.18\%&6.28&51&$>$99.99 \\ \hline
Huancayo & Polarity &0.39+0.04\%&2.21&51&$>$99.31 \\
 & Asymmetry &0.72$\pm$0.04\%&5.21&51&$>$99.99 \\ \hline
\end{tabular}
\end{table}

\end{document}